\documentclass[aps,prl,preprint,a4paper]{revtex4-1}
\usepackage{cancel}
\usepackage{amsmath}
\usepackage{amssymb}
\usepackage{amsthm}
\usepackage{graphicx}
\usepackage{dcolumn}
\usepackage{bm}
\usepackage{upgreek}
\usepackage{esint}

\usepackage{commands}

\theoremstyle{definition}

\theoremstyle{remark}

\theoremstyle{theorem}

\theoremstyle{lemma}

\theoremstyle{corollary}

\theoremstyle{proposition}

\draft 

\begin{document}

\title{Mean field theory for intense light-matter interactions in high energy density plasmas}



\author{J. W. Burby}
\affiliation{Los Alamos National Laboratory, Los Alamos, New Mexico 87545, USA}
\author{P. J. Morrison}
\affiliation{Department of Physics and Institute for Fusion Studies, The University of Texas at Austin, Austin, TX 78712, USA}


\date{\today}

\begin{abstract}
We present a generalization of Vlasov-Maxwell kinetic theory that accounts for intense electromagnetic fields. A strongly-radiating, possibly optically-thick plasma is decomposed into fragments, each comprising a charged particle together with its self-generated electromagnetic field. Assuming weak inter-fragment correlations, but strong intra-fragment correlations, a mean-field evolution equation for the single-fragment distribution functional is derived. We also identify the equation's Hamiltonian formulation. By incorporating strong correlations between a charged particle and the field it generates, the new model captures the effects of strong radiation reaction non-perturbatively. The fragment kinetic formalism offers an attractive approach to modeling exotic light-matter interactions such as nonlinear and multiple Compton scattering.
%
\end{abstract}


\maketitle 


\section{Introduction}
High-power short-pulse laser facilities currently comprise the vanguard for experimental investigations of extreme light-matter interactions. These facilities already achieve laser intensities in excess of $10^{23} \,\text{W}/\text{cm}^2$, and continually advance toward increasingly mind-bending electric field strengths. By contrast, high energy density (HED) plasma experiments like the National Ignition Facility (NIF) at Lawrence Livermore National Laboratory  and Z at Sandia National Laboratories produce relatively dim radiation, and therefore play only a modest role in pressing the boundaries of high-field science. But with the advent of burning plasma experiments at NIF, HED science will soon experience a tailwind – the field is already abuzz with chatter of NIF upgrades and a potential next-generation pulsed-power (NGPP) facility. The future holds radically new burning HED plasma experiments that promise to probe exotic radiation intensity regimes beyond the reach of state-of-the-art radiation hydrodynamic simulations. For example, extensions of the NIF DT platform to larger pure deuterium targets in \cite{Rose_2020} likely achieve radiation intensities higher than $10^{22} \,\text{W}/\text{cm}^2$. A predictive simulation capability for these plasmas should be developed before the start of next-generation experimental operations. The goal of this Article is to take a modest but crucial step in this direction by developing a radiation-free-electron interaction model suitable for aggressively-burning plasma environments in the limit of weak electron collisions. 

Modern radiation transport codes for HED plasmas model the interaction between photons and electrons as a combination of absorption, emission, and (inverse) Compton scattering. This model assumes the radiation field only slightly alters an electron’s trajectory during a single interaction event, and therefore breaks down with sufficiently intense radiation. For $1 \,\text{keV}$ photons interacting with $10 \,\text{keV}$ electrons, radiation intensities above $9\times10^{23} \,\text{W}/\text{cm}^2$ elevate the classical nonlinearity parameter \cite{Blackburn_general_2020} above $1$. Such fields strongly modify electron trajectories, leading to exotic multi-photon processes including nonlinear and double Compton scattering. Next-generation burning plasma experiments with pure deuterium fuel may probe this regime \cite{Rose_2020}. At intensities in excess of $5\times10^{26} \,\text{W}/\text{cm}^2$, and for similar particle energies, radiation drag experienced by electrons competes strongly with the usual Lorentz force, further complicating the nature of multi-photon effects \cite{Blackburn_general_2020, Koga_2005}. Such extreme intensities exist near astrophysical objects; magnetars, for example, likely experience intensities in excess of the Schwinger limit $2\times10^{29} \,\text{W}/\text{cm}^2$, at which the vacuum begins to boil. Overall, radiation transport packages in state-of-the-art multi-physics hydrocodes require entirely new physics capabilities in high-radiation-intensity regimes.  

Motivated by advances in chirped-pulse amplification, most recent research into intense light-matter interaction considers systems driven by short-pulse lasers. To model multi-photon interaction processes within a traditional PIC framework, this prior work assumes \cite{Gonoskov_2022}: (1) the radiation field decomposes into “background” laser radiation, described classically, and diffuse plasma-generated photons, described using QED, (2) the electric field is so large that the background field is constant during each interaction event – the so-called locally-constant field approximation (LCFA), and (3) the background acts like an infinite reservoir of energy and momentum during each interaction. Intense radiation in burning HED plasmas violates (1) since thermal motion provides the dominant source of photons, rather than an externally imposed laser. It also violates (2) because radiation intensity eventually decays, either spatially or temporally. Finally, (3) is inappropriate due to the crucial importance of total energy and momentum balance to macroscopic plasma dynamics. Tentative proposals for circumventing (2) or (3) either apply only to intense plane waves \cite{Heinzl_2018,Ilderton_2019} (e.g. from a laser) or require admission of complex-valued background fields \cite{Ilderton_2018}. These idiosyncrasies present serious conceptual challenges within the traditional radiation transport framework. Therefore the modeling approach adopted by the short-pulse laser community cannot be transcribed into the burning plasma context; new methods need to be developed.

\begin{figure}
\includegraphics[scale=.55]{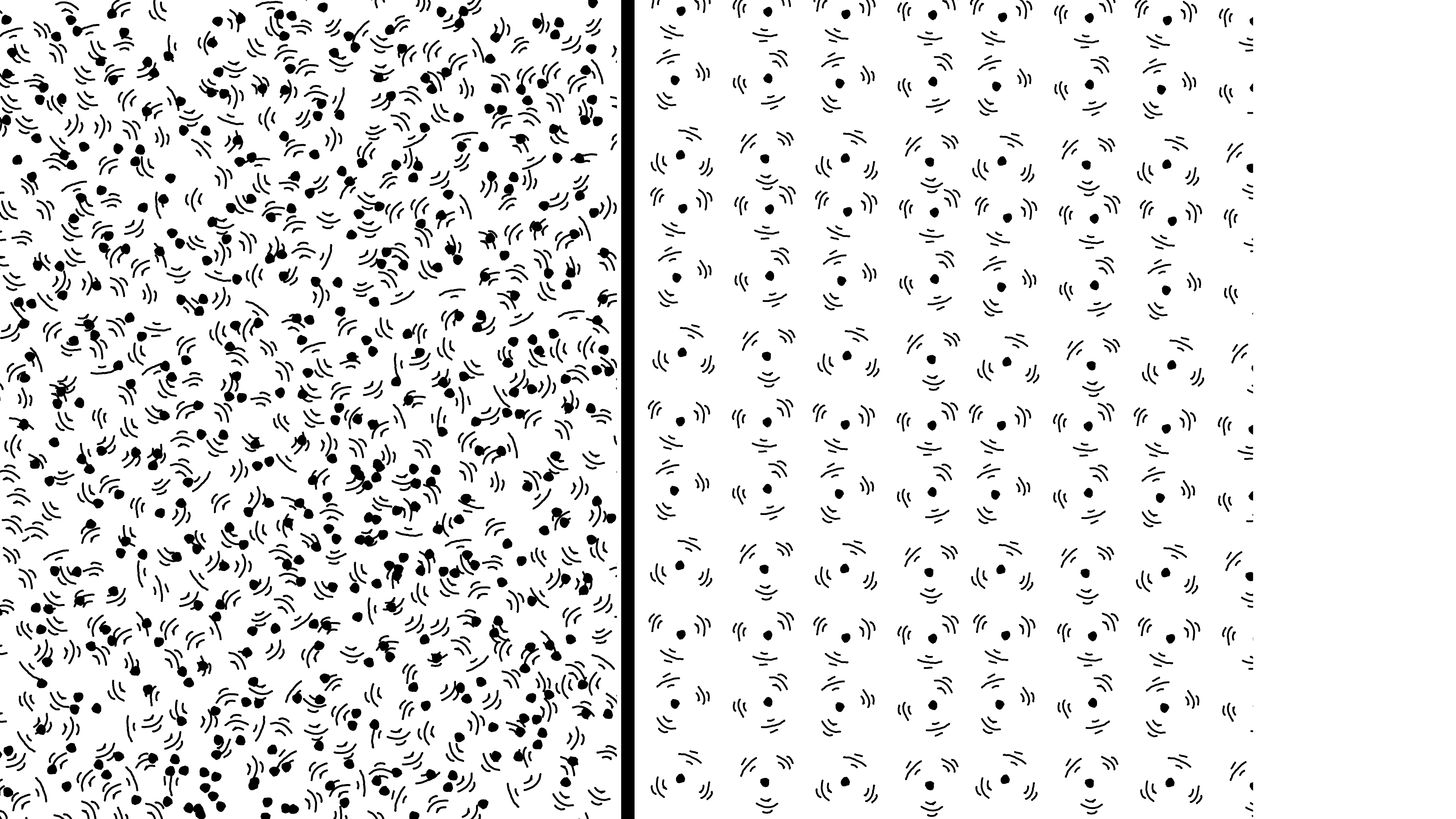}
\caption{\label{FIG1}(Left) The typical picture of a plasma; discrete particles moving through a monolithic electromagnetic field. (Right) The fragment kinetic picture; particles are grouped together with their self-produced electromagnetic field.}
\end{figure}

This Article will present a new framework for studying strong light-matter interactions in HED plasmas. The approach removes the limitations inherent to interaction models for short-pulse lasers, resulting in an electromagnetic $N$-body formalism whose complexity lies closer to that of traditional electrostatic $N$-body theory than full QED. Rather than splitting the radiation field into a strong classical background and diffuse, fully-quantum-mechanical photons, the entire intense electromagnetic field is modeled classically. The technique neglects quantum correlations between electrons, as in time-dependent Hartree theory for distinguishable particles. The electromagnetic field is expressed as a superposition of fields produced by each electron, leading to a formulation of the field-particle system as dynamics of $N$ identical interacting \textbf{fragments}. Each fragment comprises an electron together with that electron’s self-produced field. See Figure \ref{FIG1}. This model enjoys full energy- and momentum- self-consistency; no presumptive infinite reservoir of energy or momentum is necessary. Moreover, all ranges of electromagnetic field strength fit into the model seamlessly; the LCFA makes no appearance. In particular, due to the non-perturbative inclusion of the electromagnetic self-force for each fragment, the particle model accommodates both Lorentz-force-dominated dynamics at modest field strengths and (classical) radiation-reaction-dominated dynamics at high field strengths \cite{Blackburn_general_2020}. Pair production is not included. Both the $N$-fragment model and its mean-field limit will be presented. The mean field limit neglects collisions between fragments. Equivalently, it neglects collisions between electrons while maintaining strong radiation-electron correlations. Future work will identify the fragment-fragment collision operator.

\section{notation}
We make use of the following notation repeatedly in the text. Fragment labels are denoted $z_0 = (\bm{x}_0,\bm{v}_0,\bm{e}_0,\bm{b}_0)$. The distribution of fragment labels is $\mathcal{F}_0$. Points in fragment phase space are denoted $z = (\bm{x}_0,\bm{v}_0,\bm{e}_0,\bm{b}_0)$. The distribution of fragments is $\mathcal{F}$. The Lagrangian configuration map that maps fragment labels to their corresponding fragments at the current time is denoted $Z = (\bm{X},\bm{V},\bm{\mathcal{E}},\bm{\mathcal{B}})$. The collective electric and magnetic fields are $\bm{E}\text{ and }\bm{B}$, respectively. Individual fragments are denoted $Z_i = (\bm{X}_i,\bm{V}_i,\bm{\mathcal{E}}_i,\bm{\mathcal{B}}_i)$.

\section{Molecular chaos redux}
Recall the statistical argument used to deduce the Vlasov-Poisson system from the electrostatic $N$-body problem. A collection of $N\gg 1$ charged particles, for simplicity each with charge $m$ and mass $q$, move according to the equations of motion
\begin{align*}
\dot{\bm{X}}_i = \bm{V}_i,\quad m\,\dot{\bm{V}}_i=\bm{F}_i(\bm{X}_i),
\end{align*}
where the field that determines the force on the $i^{\text{th}}$ particle is 
\begin{align*}
\bm{F}_i(\bm{x})=\sum_{j\neq i}\frac{q^2}{4\pi\epsilon_0}\frac{\bm{x}-\bm{X}_j}{|\bm{x}-\bm{X}_j|^3}.
\end{align*}
Under the mild assumption of random initial conditions for each particle, the field $\bm{F}_i$ is itself a random variable. Long ago, Boltzmann hypothesized that if the particle initial conditions are independent and identically distributed, molecular chaos will ensure that the particle states remain independent and identically distributed as time advances. In this scenario, the Law of Large Numbers says the field $\bm{F}_i$ is given very nearly by the non-random expression
\begin{align*}
\bm{F}_i(\bm{x})\approx \int \frac{q^2}{4\pi\epsilon_0}\frac{\bm{x}-\overline{\bm{x}}}{|\bm{x}-\overline{\bm{x}}|^3}\,f(\overline{\bm{x}},\overline{\bm{v}})\,d^3\overline{\bm{x}}\,d^3\overline{\bm{v}},
\end{align*}
where $f(\bm{x},\bm{v})$ denotes the single-particle (Vlasov) distribution function. Note that $f$ does not depend on $j$; this is an easy consequence of the statistical indistinguishability of particles at $t=0$. Boltzmann's molecular chaos hypothesis therefore implies the particle equations of motion approximately simplify to
\begin{align*}
\dot{\bm{X}}_i = \bm{V}_i,\quad m\,\dot{\bm{V}}_i = \int \frac{q^2}{4\pi\epsilon_0}\frac{\bm{X}_i-\overline{\bm{x}}}{|\bm{X}_i-\overline{\bm{x}}|^3}\,f(\overline{\bm{x}},\overline{\bm{v}})\,d^3\overline{\bm{x}}\,d^3\overline{\bm{v}}.
\end{align*}
The right-hand-side of the momentum equation for the $i^{\text{th}}$ particle only depends on $\bm{X}_i$ and the single-particle distribution function in this limit. The evolution equation for $f$ therefore closes on itself according to
\begin{align*}
\dot{f} + \nabla\cdot(\bm{v}\,f) + \partial_{\bm{v}}\cdot\left(m^{-1}\left[\int \frac{q^2}{4\pi\epsilon_0}\frac{\bm{x}-\overline{\bm{x}}}{|\bm{x}-\overline{\bm{x}}|^3}\,f(\overline{\bm{x}},\overline{\bm{v}})\,d^3\overline{\bm{x}}\,d^3\overline{\bm{v}}\right]\,f\right) = 0.
\end{align*}
This quadratically-nonlinear evolution equation for $f$ recovers the single-species Vlasov-Poisson system for a repulsive interaction. 

The preceding argument generalizes to the electromagnetic $N$-body problem. The generalized argument produces a fundamental extension of the famous Vlasov-Maxwell system that accounts for the infamous electromagnetic self force in a natural way. Rather than treating the radiated field as a dissipative drain of energy, the new model accounts for the energetics (and momentum transport) of radiation reaction self-consistently. The results reported here therefore represent a crucial ingredient in the evolving theoretical understanding of plasmas that interact with high-intensity laser fields. They also offer a novel first-principles approach to modeling radiation transport in strongly-radiating plasmas.

%

\section{Formulation of the electromagnetic $N$-body problem}
The electromagnetic $N$-body problem suffers from several complications that do not plague the electrostatic problem. The first of these relates to the well-known mathematical difficulties surrounding the electromagnetic self-force. What equations of motion do the charged particles obey? This seemingly innocuous question continues to fuel a lively discourse in the physics community owing to the fact that the electromagnetic field produced by a moving point charge assumes an infinite value at the particle's location. 

Abraham, and later Dirac, famously studied this problem in the special case $N=1$. The equations of motion they derived emerge from analyses involving classical mass renormalization and ``integrating out" the electromagnetic field. The Lorentz-Abraham-Dirac (LAD) form of the radiation drag force, or a regularization due to Landau and Lifshitz, is frequently used in PIC simulations involving high-energy electrons \cite{Vranic_2016}. Unfortunately, the LAD result breaks down as soon as reabsorption of the radiated field plays a physically-significant role, as would be true in  optically-thick plasmas. This means the LAD equations of motion cannot be used to formulate a realistic electromagnetic $N$-body problem in optically-thick plasmas. Restricting attention to only optically-thin plasmas rules out many interesting high-energy-density environments, and so a better theory is necessary. 

Feynman-Wheeler absorber theory provides a different approach specifically formulated to allow for an arbitrary number of particles. However, absorber theory appears to be poorly suited to numerical simulations since both the advanced and retarded electromagnetic Greens functions play a fundamental role in the formalism. Moreover, the theory somewhat implausibly predicts that each particle's self-force vanishes. While the Feynman-Wheeler approach may ultimately interact nicely with the molecular chaos hypothesis, a simpler approach seems desirable.

This Article will sidestep subtleties associated with point-like particles by restricting attention to finite-sized non-relativistic particles. There are several reasons for doing so. First, all serious mathematical complications arising from point particles disappear when each particle enjoys a spatially-extended charge distribution; the electromagnetic field remains finite everywhere, and the force on each particle is given by the standard volumetric Lorentz formula. Second, by considering particles that move at non-relativistic speeds, standard rigid body theory provides a plausible method to model each particle's degrees of freedom. Finally, 
the statistical analysis of finite-sized particles readily generalizes to account for all of the effects of special relativity upon replacing the rigid bodies with classical Dirac or Klein-Gordon fields. Thus, there is really no loss of generality in presenting its crucial statistical argument in the rigid body framework.

The electromagnetic $N$-body model for finite-sized particles of mass $m$ and charge $q$ asserts that the electromagnetic field $(\bm{E},\bm{B})$ and the particle phase space locations $(\bm{X}_i,\bm{V}_i)$ evolve according to the evolution equations
\begin{align}
\dot{\bm{X}}_i & = \bm{V}_i\label{em_nbody_1}\\
m\,\dot{\bm{V}}_i &= q\,\int (\bm{E}(\bm{r}) + \bm{V}_i\times\bm{B}(\bm{r}))\,\mathcal{W}(\bm{r}-\bm{X}_i)\,d^3\bm{r}\label{em_nbody_2}\\
\epsilon_0\,\dot{\bm{E}}(\bm{r}) & = \mu_0^{-1}\nabla\times\bm{B}(\bm{r}) - \sum_{i} q\,\bm{V}_i\,\mathcal{W}(\bm{r}-\bm{X}_i) \\
\dot{\bm{B}}(\bm{r}) & = -\nabla\times\bm{E}(\bm{r}).\label{em_nbody_4}
\end{align}
Here the shape function $\mathcal{W}(\bm{\xi})$ represents the normalized charge distribution of each particle; it has compact support and unit total mass, $\int \mathcal{W}(\bm{\xi})\,d^3\bm{\xi} = 1$. This model easily generalizes to include multiple particle species.

\section{Identical subsystems}
The statistical argument that leads from the electrostatic $N$-body problem to the Vlasov-Poisson system relies crucially on decomposing the many-body system into $N$ interacting identical subsystems. Each subsystem is given by one of the particles. In the electromagnetic case, after decomposing by particles, the electromagnetic field remains, thus spoiling homogeneity of the decomposition. If the electromagnetic problem does admit some homogeneous subsystem decomposition, that decomposition must therefore differ from the electrostatic decomposition.

To find a homogeneous subsystem decomposition of the electromagnetic $N$-body problem, first express the electromagnetic field $(\bm{E},\bm{B})$ as the superposition of fields $(\bm{\mathcal{E}}_i,\bm{\mathcal{B}}_i)$ produced by each particle,
\begin{align}
\bm{E} = \sum_{i} \bm{\mathcal{E}}_i,\quad \bm{B} = \sum_{i}\bm{\mathcal{B}}_i.\label{superposition}
\end{align}
Now group each particle $(\bm{X}_i,\bm{V}_i)$ together with the field that particle produces $(\bm{\mathcal{E}}_i,\bm{\mathcal{B}}_i)$ to form the corresponding \textbf{fragment} $Z_i = (\bm{X}_i,\bm{V}_i,\bm{\mathcal{E}}_i,\bm{\mathcal{B}}_i)$. The following argument shows that the electromagnetic $N$-body problem decomposes into $N$ interacting identical fragments.

Because $(\bm{\mathcal{E}}_i,\bm{\mathcal{B}}_i)$ is the field produced by $(\bm{X}_i,\bm{V}_i)$ it must satisfy Maxwell's equations with sources determined by $(\bm{X}_i,\bm{V}_i)$ alone. Each fragment therefore satisfies the system of evolution equations
\begin{align}
\dot{\bm{X}}_i & = \bm{V}_i\label{em_split_1}\\
m\,\dot{\bm{V}}_i &= q\,\int (\bm{E}(\bm{r}) + \bm{V}_i\times\bm{B}(\bm{r}))\,\mathcal{W}(\bm{r}-\bm{X}_i)\,d^3\bm{r}\label{em_split_2}\\
\epsilon_0\,\dot{\bm{\mathcal{E}}}_i(\bm{r}) & = \mu_0^{-1}\nabla\times\bm{\mathcal{B}}_i(\bm{r}) - q\,\bm{V}_i\,\mathcal{W}(\bm{r}-\bm{X}_i)\label{em_split_3}\\
\dot{\bm{\mathcal{B}}}_i(\bm{r}) & = -\nabla\times\bm{\mathcal{E}}_i(\bm{r}),\label{em_split_4}
\end{align}
where $\bm{E}$ and $\bm{B}$ are given by \eqref{superposition}.
While the phase space for the original formulation of the electromagnetic $N$-body problem contained tuples of the form $(\bm{X}_1,\bm{V}_1,\dots,\bm{X}_N,\bm{V}_N,\bm{E},\bm{B})$, the phase space for this new system comprises tuples of the form $(Z_1,\dots,Z_N)$. In this sense \eqref{em_split_1}-\eqref{em_split_4} is ``larger" than \eqref{em_nbody_1}-\eqref{em_nbody_4}. On the other hand, if $(Z_1,\dots,Z_N)$ is a solution of \eqref{em_split_1}-\eqref{em_split_4} then $(\bm{X}_1,\bm{V}_1,\dots,\bm{X}_N,\bm{V}_N,\bm{E},\bm{B})$, with $(\bm{E},\bm{B})$ given by \eqref{superposition}, is a solution of \eqref{em_nbody_1}-\eqref{em_nbody_4}. In fact, all solutions of \eqref{em_nbody_1}-\eqref{em_nbody_4} arise in this manner. The larger system \eqref{em_split_1}-\eqref{em_split_4} therefore provides a physically-valid reformulation of the electromagnetic $N$-body problem. Moreover, the new formulation manifestly admits a homogeneous decomposition into subsystems given by the fragments. In this new form, the electromagnetic $N$-body problem is amenable to application of Boltzmann's molecular chaos hypothesis.

\section{The generalized molecular chaos hypothesis}
Boltzmann hypothesized that statistical independence among interacting particles propagates through time. Boltzmann's prediction is sometimes referred to as \textbf{propagation of chaos}. Chaos presumably also propagates among particles in the electromagnetic $N$-body problem. But the electromagnetic field must also play an interesting role in this statistical phenomenon since each particle is strongly correlated with the field it produces. In other words, the electromagnetic $N$-body problem motivates refining Boltzmann's hypothesis so as to account for the electromagnetic field.

At face value, Boltzmann's hypothesis pertains to statistical properties of particles. At a more fundamental level, the particles carry no fundamental importance \emph{per se}. Instead Boltzmann's hypothesis may be understood as a statement concerning the identical subsystems that comprise the electrostatic $N$-body problem; it just so happens that those subsystems coincide with particles for the electrostatic problem. From this perspective, the most meaningful way to formulate Boltzmann's hypothesis for the electromagnetic problem is the following.
\begin{itemize}
\item \textbf{Generalized propagation of chaos:} If the fragments $Z_i$ are independent and identically distributed initially then they will remain so for all time.
\end{itemize}
In other words, ``particles" in the electrostatic formulation should be replaced with ``fragments" in the electromagnetic formulation. Crucially, generalized propagation of chaos does not assert that correlations between a particle and its own field vanish. Indeed, the various components of each $Z_i$ may exhibit strong correlations among themselves without suffering correlations with $Z_j$, $j\neq i$. 

Intra-fragment correlations in general, and particle-self-field correlations in particular, cannot be ignored when single-particle radiation reaction plays an important dynamical role. For if the correlation between a particle and its own field were zero the mean force felt by the particle would not include radiation drag, but only that force associated with the smooth collective electromagnetic field. 


\section{Derivation of the fragment kinetic equation}

Now consider a collection of $N\gg 1$ fragments, for simplicity each with charge $m$ and mass $q$, moving according to the equations of motion \eqref{em_split_1}-\eqref{em_split_4}. Observe that these equations have the general form
\begin{align*}
\dot{Z}_i = \mathcal{U}_i(Z_i;Z_{j\neq i}),\quad i=1,\dots,N.
\end{align*}
Under the mild assumption of random initial conditions for each fragment, the field $\mathcal{U}_i(z;Z_{j\neq i})$, where $z = (\bm{x},\bm{v},\bm{e},\bm{b})$ is an arbitrary element of the single-fragment phase space, is itself a random variable. Also assuming validity of generalized propagation of chaos, the Law of Large Numbers says $\mathcal{U}_i(z;Z_{j\neq i})$ is given very nearly by the non-random expression $\mathcal{U}_i(z;Z_{j\neq i})\approx {\mathcal{U}}(z)$. The components of $\mathcal{U}(z) = (\mathcal{U}^{\bm{x}}(z),\mathcal{U}^{\bm{v}}(z),\mathcal{U}^{\bm{e}}(z),\mathcal{U}^{\bm{b}}(z))$ are given by
\begin{align}
\mathcal{U}^{\bm{x}}(z)& = \bm{v}\label{velocity_1}\\
\mathcal{U}^{\bm{v}}(z)& = \frac{q}{m}\,\int (\bm{e}(\bm{r}) + \bm{v}\times\bm{b}(\bm{r}))\,\mathcal{W}(\bm{r}-\bm{x})\,d^3\bm{r}\nonumber\\
&+[1-N^{-1}]\frac{q}{m}\,\int ( \bm{{E}}(\bm{r}) + \bm{v}\times\bm{{B}}(\bm{r}))\,\mathcal{W}(\bm{r}-\bm{x})\,d^3\bm{r}\\
\mathcal{U}^{\bm{e}}(z)& =  \mu_0^{-1}\nabla\times\bm{b}(\bm{r}) - q\,\bm{v}\,\mathcal{W}(\bm{r}-\bm{x})\\
\mathcal{U}^{\bm{b}}(z) & = -\nabla\times\bm{e}(\bm{r}).\label{velocity_4}
\end{align}
By a slight abuse of notation, here $(\bm{{E}},\bm{{B}}) = \int (\bm{e},\bm{b})\,\mathcal{F}(\bm{x},\bm{v},\bm{e},\bm{b})\,d^3\bm{x}\,d^3\bm{v}\,\mathcal{D}\bm{e}\,\mathcal{D}\bm{b}$, where $\mathcal{F}$ is the single-fragment (Vlasov) distribution functional. (Beware: $\mathcal{F}\,d^3\bm{x}\,d^3\bm{v}\,\mathcal{D}\bm{e}\,\mathcal{D}\bm{b}$ must be interpreted as a probability measure on the infinite-dimensional single-fragment phase space of tuples $(\bm{x},\bm{v},\bm{e},\bm{b})$, and \emph{not} a measure on the finite-dimensional space $\mathbb{R}^3\times\mathbb{R}^3\times\mathbb{R}^3\times\mathbb{R}^3$.)
Generalized propagation of chaos therefore implies the fragment equations of motion approximately simplify to

\begin{align*}
\dot{\bm{X}}_i& = \bm{V}_i\\
\dot{\bm{V}}_i& = \frac{q}{m}\,\int (\bm{\mathcal{E}}_i(\bm{r}) + \bm{V}_i\times\bm{\mathcal{B}}(\bm{r}))\,\mathcal{W}(\bm{r}-\bm{X}_i)\,d^3\bm{r}\\
&+[1-N^{-1}]\frac{q}{m}\,\int ( \bm{{E}}(\bm{r}) + \bm{V}\times\bm{{B}}(\bm{r}))\,\mathcal{W}(\bm{r}-\bm{X}_i)\,d^3\bm{r}\\
\dot{\bm{\mathcal{E}}}_i(\bm{r})& =  \mu_0^{-1}\nabla\times\bm{\mathcal{B}}_i(\bm{r}) - q\,\bm{V}_i\,\mathcal{W}(\bm{r}-\bm{X}_i)\\
\dot{\bm{\mathcal{B}}}_i(\bm{r}) & = -\nabla\times\bm{\mathcal{E}}_i(\bm{r}).
\end{align*}
The right-hand-sides of these equations depend only on $Z_i$ and the single-fragment distribution function in this limit. The evolution equation for $\mathcal{F}$ therefore closes on itself according to
\begin{align}
\dot{\mathcal{F}}\,d^3\bm{x}\,d^3\bm{v}\,\mathcal{D}\bm{e}\,\mathcal{D}\bm{b} + \mathcal{L}_{\mathcal{U}}({\mathcal{F}}\,d^3\bm{x}\,d^3\bm{v}\,\mathcal{D}\bm{e}\,\mathcal{D}\bm{b}) = 0,\label{functional_vlasov}
\end{align}
where $\mathcal{U}$ is the vector field on single-fragment phase space given as above and $\mathcal{L}_{\mathcal{U}}$ denotes the Lie derivative of measures along $\mathcal{U}$.
The quadratically-nonlinear evolution equation \eqref{functional_vlasov} defines the mean-field limit of the fragment kinetic equation. It plays a role in electromagnetic $N$-body dynamics analogous to the Vlasov-Poisson equation in electrostatic $N$-body dynamics. While the mean-field model fails to include collisions between fragments, it retains strong correlations between each electron and the electromagnetic field the electron produces.

\section{Variational and Hamiltonian formulation of the fragment kinetic equation}
Like the collisionless Vlasov-Poisson system, the mean-field fragment kinetic equation \eqref{functional_vlasov} satisfies a variational principle. Also like the Vlasov-Poisson case, the natural variational principle for the fragment kinetic equation makes use of Lagrangian, as opposed to Eulerian, variables. And like the Vlasov-Maxwell variation principle, the fragment variational principle uses the vector potential (in temporal gauge) instead of the magnetic field.  

The Lagrangian configuration map for the fragment kinetic equation comprises a time-dependent invertible mapping $Z:z_0\mapsto Z(z_0)=(\bm{X}(z_0),\bm{V}(z_0),\bm{\mathcal{E}}(z_0),\bm{\mathcal{A}}(z_0))$ from \textbf{fragment labels} $z_0$ into the gauge-dependent single-fragment phase space. In terms of the Lagrangian configuration map, the phase space Lagrangian for the fragment kinetic equation is
\begin{align}
&L(Z,\dot{Z})  = \int \bigg(m\,\bm{V} + q\int \bm{\mathcal{A}}(\bm{r})\,\mathcal{W}(\bm{r}-\bm{X})\,d^3\bm{r}\bigg)\cdot \dot{\bm{X}}\,\mathcal{F}_{0}\,d^3\bm{x}_0\,d^3\bm{v}_0\,\mathcal{D}\bm{e}_0\,\mathcal{D}\bm{a}_0\nonumber\\
&-\epsilon_0\int\bigg(\int \bm{\mathcal{E}}\cdot\dot{\bm{\mathcal{A}}}\,d^3\bm{r}\bigg)\,\mathcal{F}_0\,d^3\bm{x}_0\,d^3\bm{v}_0\,\mathcal{D}\bm{e}_0\,\mathcal{D}\bm{a}_0\nonumber\\ 
&+ [1-N^{-1}]\int \bigg( q\int \bm{A}(\bm{r})\,\mathcal{W}(\bm{r}-\bm{X})\,d^3\bm{r}\bigg)\cdot \dot{\bm{X}}\,\mathcal{F}_{0}\,d^3\bm{x}_0\,d^3\bm{v}_0\,\mathcal{D}\bm{e}_0\,\mathcal{D}\bm{a}_0\nonumber\\
&- [1-N^{-1}]\epsilon_0\int \bm{E}\cdot \dot{\bm{A}}\,d^3\bm{r} - H(Z),
\end{align}
where the Hamiltonian ${H}$ is 
\begin{align}
\mathcal{H}(Z) &= \int \frac{1}{2}m|\bm{V}|^2\,\mathcal{F}_0\,d^3\bm{x}_0\,d^3\bm{v}_0\,\mathcal{D}\bm{e}_0\,\mathcal{D}\bm{a}_0 \nonumber\\
&+ \int\bigg(\int \frac{1}{2}\epsilon_0\,|\bm{\mathcal{E}}|^2\,d^3\bm{r}+ \int\int \frac{1}{2}\mu_0^{-1}|\nabla\times\bm{\mathcal{A}}|^2\,d^3\bm{r}\,\bigg)\mathcal{F}_0\,d^3\bm{x}_0\,d^3\bm{v}_0\,\mathcal{D}\bm{e}_0\,\mathcal{D}\bm{a}_0 \nonumber\\
& +[1-N^{-1}] \int \frac{1}{2}\epsilon_0\,|\bm{E}|^2\,d^3\bm{r} +[1-N^{-1}] \int\frac{1}{2}\mu_0^{-1}|\nabla\times\bm{A}|^2\,d^3\bm{r},
\end{align}
and $\mathcal{F}_0$ denotes the \textbf{fragment label distribution}. In this variational formulation $\mathcal{F}_0$ is a time-independent parameter, in terms of which the fragment distribution is defined according to
\begin{align*}
Z_*(\mathcal{F}_0\,d^3\bm{x}_0\,d^3\bm{v}_0\,\mathcal{D}\bm{e}_0\,\mathcal{D}\bm{a}_0) = \mathcal{F}\,d^3\bm{x}\,d^3\bm{v}\,\mathcal{D}\bm{e}\,\mathcal{D}\bm{a},
\end{align*}
where $Z_*$ denotes the pushforward of measures along the mapping $Z$.

Modulo total time derivatives, the first variation of the phase space Lagrangian with respect to $Z$ is given by
\begin{align*}
&\delta_{\bm{X}}L =-\int \bigg(m\,\dot{\bm{V}} + q\int \dot{\bm{\mathcal{A}}}(\bm{r})\,\mathcal{W}(\bm{r}-\bm{X})\,d^3\bm{r}\bigg)\cdot \delta{\bm{X}}\,\mathcal{F}_{0}\,d^3\bm{x}_0\,d^3\bm{v}_0\,\mathcal{D}\bm{e}_0\,\mathcal{D}\bm{a}_0\\
&+\int \bigg( q\int \dot{\bm{X}}\times\bm{\mathcal{B}}(\bm{r})\,\mathcal{W}(\bm{r}-\bm{X})\,d^3\bm{r}\bigg)\cdot \delta\bm{X}\,\mathcal{F}_{0}\,d^3\bm{x}_0\,d^3\bm{v}_0\,\mathcal{D}\bm{e}_0\,\mathcal{D}\bm{a}_0\\
& -[1-N^{-1}]\int \bigg( q\int \dot{\bm{A}}(\bm{r})\,\mathcal{W}(\bm{r}-\bm{X})\,d^3\bm{r}\bigg)\cdot \delta{\bm{X}}\,\mathcal{F}_{0}\,d^3\bm{x}_0\,d^3\bm{v}_0\,\mathcal{D}\bm{e}_0\,\mathcal{D}\bm{a}_0\\
&+[1-N^{-1}]\int \bigg( q\int \dot{\bm{X}}\times\bm{{B}}(\bm{r})\,\mathcal{W}(\bm{r}-\bm{X})\,d^3\bm{r}\bigg)\cdot \delta\bm{X}\,\mathcal{F}_{0}\,d^3\bm{x}_0\,d^3\bm{v}_0\,\mathcal{D}\bm{e}_0\,\mathcal{D}\bm{a}_0\\
& - \int \frac{\delta H}{\delta\bm{X}}\cdot \delta\bm{X}\,\mathcal{F}_{0}\,d^3\bm{x}_0\,d^3\bm{v}_0\,\mathcal{D}\bm{e}_0\,\mathcal{D}\bm{a}_0\\
&\delta_{\bm{V}}L = \int \bigg(m\,\dot{\bm{X}} - \frac{\delta H}{\delta\bm{V}} \bigg)\cdot \delta{\bm{V}}\,\mathcal{F}_{0}\,d^3\bm{x}_0\,d^3\bm{v}_0\,\mathcal{D}\bm{e}_0\,\mathcal{D}\bm{a}_0\\
&\delta_{\bm{\mathcal{E}}}L =  -\epsilon_0\int\bigg(\int \dot{\bm{\mathcal{A}}}\cdot \delta\bm{\mathcal{E}}\,d^3\bm{r}\bigg)\,\mathcal{F}_0\,d^3\bm{x}_0\,d^3\bm{v}_0\,\mathcal{D}\bm{e}_0\,\mathcal{D}\bm{a}_0\\
&- [1-N^{-1}]\epsilon_0\int\bigg(\int \dot{\bm{A}}\cdot \delta\bm{\mathcal{E}}\,d^3\bm{r}\bigg)\,\mathcal{F}_0\,d^3\bm{x}_0\,d^3\bm{v}_0\,\mathcal{D}\bm{e}_0\,\mathcal{D}\bm{a}_0\\
& - \int \bigg(\int \frac{\delta H}{\delta \bm{\mathcal{E}}}\cdot \delta\bm{\mathcal{E}}\,d^3\bm{r}\bigg)\,\mathcal{F}_0\,d^3\bm{x}_0\,d^3\bm{v}_0\,\mathcal{D}\bm{e}_0\,\mathcal{D}\bm{a}_0\\
&\delta_{\bm{\mathcal{A}}}L = \int \bigg(\int q\,\dot{\bm{X}}\,\mathcal{W}(\bm{r}-\bm{X})\cdot \delta\bm{\mathcal{A}}\,d^3\bm{r}\bigg)\,\mathcal{F}_0\,d^3\bm{x}_0\,d^3\bm{v}_0\,\mathcal{D}\bm{e}_0\,\mathcal{D}\bm{a}_0\\
&+\epsilon_0\int\bigg(\int \dot{\bm{\mathcal{E}}}\cdot\delta{\bm{\mathcal{A}}}\,d^3\bm{r}\bigg)\,\mathcal{F}_0\,d^3\bm{x}_0\,d^3\bm{v}_0\,\mathcal{D}\bm{e}_0\,\mathcal{D}\bm{a}_0\nonumber\\ 
&+ [1-N^{-1}]\int \bigg(\int \left(\int q\,\dot{\bm{X}}\,\mathcal{W}(\bm{r}-\bm{X})\,\mathcal{F}_0\,d^3\bm{x}_0\,d^3\bm{v}_0\,\mathcal{D}\bm{e}_0\,\mathcal{D}\bm{a}_0\right)\cdot \delta\bm{\mathcal{A}}\,d^3\bm{r}\bigg)\,\mathcal{F}_0\,d^3\bm{x}_0\,d^3\bm{v}_0\,\mathcal{D}\bm{e}_0\,\mathcal{D}\bm{a}_0\\
&+ [1-N^{-1}]\epsilon_0\int\bigg(\int \dot{\bm{E}}\cdot \delta{\bm{\mathcal{A}}}\,d^3\bm{r}\bigg)\,\mathcal{F}_0\,d^3\bm{x}_0\,d^3\bm{v}_0\,\mathcal{D}\bm{e}_0\,\mathcal{D}\bm{a}_0\\
& - \int\bigg(\int \frac{\delta H}{\delta\bm{\mathcal{A}}}\cdot \delta\bm{\mathcal{A}}\bigg)\,\mathcal{F}_0\,d^3\bm{x}_0\,d^3\bm{v}_0\,\mathcal{D}\bm{e}_0\,\mathcal{D}\bm{a}_0.
\end{align*}
The Euler-Lagrange equations are therefore
\begin{align*}
\dot{\bm{X}} & = \frac{1}{m}\,\frac{\delta H}{\delta \bm{V}}\\
m\,\dot{\bm{V}} & = -\frac{\delta H}{\delta\bm{X}}-q\int \left(\dot{\bm{\mathcal{A}}} + [1-N^{-1}]\dot{\bm{A}}\right)(\bm{r})\,\mathcal{W}(\bm{r}-\bm{X})\,d^3\bm{r}\\
& + q\int\left( \dot{\bm{X}}\times (\bm{\mathcal{B}} + [1-N^{-1}]\bm{B})(\bm{r})\right)\,\mathcal{W}(\bm{r}-\bm{X})\,d^3\bm{r}\\
\dot{\bm{\mathcal{A}}} + [1-N^{-1}]\dot{\bm{A}} & = -\frac{1}{\epsilon_0}\frac{\delta H}{\delta\bm{\mathcal{E}}}\\
\epsilon_0(\dot{\bm{\mathcal{E}}} + [1-N^{-1}]\dot{\bm{E}}) & = \frac{\delta H}{\delta\bm{\mathcal{A}}} - q\,\dot{\bm{X}}\,\mathcal{W}(\bm{r}-\bm{X})\\
& - [1-N^{-1}]\int q\,\dot{\bm{X}}\,\mathcal{W}(\bm{r}-\bm{X})\,\mathcal{F}_0\,d^3\bm{x}_0\,d^3\bm{v}_0\,\mathcal{D}\bm{e}_0\,\mathcal{D}\bm{a}_0
\end{align*}
which imply the evolution equations


\begin{align}
\dot{\bm{X}} & = \frac{1}{m}\,\frac{\delta H}{\delta \bm{V}}\label{ham_gen_1}\\
\dot{\bm{V}} & =  -\frac{1}{m}\frac{\delta H}{\delta\bm{X}}+\frac{q}{m}\int \left(\frac{1}{\epsilon_0}\frac{\delta H}{\delta\bm{\mathcal{E}}}\right)(\bm{r})\,\mathcal{W}(\bm{r}-\bm{X})\,d^3\bm{r}\nonumber\\
& + \frac{q}{m}\int\left( \frac{1}{m}\,\frac{\delta H}{\delta \bm{V}}\times (\bm{\mathcal{B}} + [1-N^{-1}]\bm{B})(\bm{r})\right)\,\mathcal{W}(\bm{r}-\bm{X})\,d^3\bm{r}\\
\dot{\bm{\mathcal{A}}}  & = -\frac{1}{\epsilon_0}\frac{\delta H}{\delta\bm{\mathcal{E}}}+ [1-N^{-1}]\left(N^{-1}\int \frac{1}{\epsilon_0}\frac{\delta H}{\delta\bm{\mathcal{E}}}\,\mathcal{F}_0\,d^3\bm{x}_0\,d^3\bm{v}_0\,\mathcal{D}\bm{e}_0\,\mathcal{D}\bm{a}_0\right)\\
\epsilon_0\,\dot{\bm{\mathcal{E}}}  & = \frac{\delta H}{\delta\bm{\mathcal{A}}} -[1-N^{-1}]\bigg(N^{-1}\int\frac{\delta H}{\delta\bm{\mathcal{A}}}\,\mathcal{F}_0\,d^3\bm{x}_0\,d^3\bm{v}_0\,\mathcal{D}\bm{e}_0\,\mathcal{D}\bm{a}_0\bigg)\nonumber\\
& - \frac{q}{m}\,\frac{\delta H}{\delta \bm{V}}\,\mathcal{W}(\bm{r}-\bm{X}).\label{ham_gen_4}
\end{align}
According to the functional derivative formulas
\begin{gather*}
\frac{\delta H}{\delta\bm{X}}  = 0,\quad  \frac{\delta H}{\delta\bm{V}}  = m\,\bm{V},\\
 \frac{\delta H}{\delta\bm{\mathcal{E}}}  = \epsilon_0\,(\bm{\mathcal{E}} + [1-N^{-1}]\bm{E}),\quad \frac{\delta H}{\delta\bm{\mathcal{A}}}  = \mu_0^{-1}(\bm{\mathcal{B}} + [1-N^{-1}]\bm{B}),
\end{gather*}
the evolution equations \eqref{ham_gen_1}-\eqref{ham_gen_4} simplify to $\dot{Z} = \mathcal{U}(Z)$, with $\mathcal{U}$ given by \eqref{velocity_1}-\eqref{velocity_4}.

\section{A Galerkin truncation of the fragment kinetic equation}
Both the well-known Vlasov-Poisson system and the fragment kinetic system comprise nonlinear advection equations. However, while Vlasov-Poisson concerns advection of probability on a finite-dimensional space, in the fragment kinetic context the unknown probability is defined on an infinite-dimensional space. In order to convey the physical content of the latter dichotomy, we will now retrace the derivation of the fragment kinetic equation while suppressing all but a finite number of degrees of freedom for the electromagnetic field of each fragment. This will result in a truncation of the fragment kinetic equation that comprises an integro-differential equation of the usual sort, rather than a functional integro-differential equation. While this truncation does not retain the full physical fidelity of the complete fragment kinetic equation, it nevertheless serves to illuminate the essential differences with the traditional Vlasov-Maxwell formalism.

Suppose that the electromagnetic field of each fragment takes the special form
\begin{align}
\bm{a}  = a_x\,\psi(y)\,\hat{\bm{x}} +a_y\, \chi^\prime(y)\,\hat{\bm{y}},\quad \bm{e} & = e_x\,\psi(y)\,\hat{\bm{x}} +e_y\,\chi^\prime(y)\,\hat{\bm{y}},\label{simple_fields}
\end{align}
where $\psi(y),\chi(y)$ are fixed smooth single-variable functions and $a_x,a_y,e_x,e_y\in\mathbb{R}$ represent the dynamical degrees of freedom for the field in this reduced description. For fields of this form the (gauge-dependent) phase space location of a single fragment reduces to the tuple $z= (\bm{r},\bm{v},a_x,a_y,e_x,e_y)\in\mathbb{R}^3\times\mathbb{R}^3\times\mathbb{R}^4\equiv \mathcal{P}_f$, and the single-fragment distribution functional becomes merely a non-negative function $\mathcal{F} = \mathcal{F}(\bm{r},\bm{v},a_x,a_y,e_x,e_y)$ on the finite-dimensional space $\mathcal{P}_f$. Accordingly, the fragment kinetic phase space Lagrangian simplifies according to
\begin{align}
L(Z,\dot{Z}) & = m\,\langle V_x \dot{X}\rangle +m\,\langle V_y\,\dot{Y}\rangle+m\,\langle V_z\,\dot{Z}\rangle+ q\,\langle \mathcal{A}_x\,\psi(Y)\,\dot{X}\rangle + q\,\langle \mathcal{A}_y\,\chi^\prime(Y)\,\dot{Y}\rangle
\nonumber\\
&-\epsilon_0\bigg( M_{\psi}\,\langle\mathcal{E}_x\,\dot{\mathcal{A}}_x\rangle +M_{\chi^\prime}\,\langle \mathcal{E}_y\,\dot{\mathcal{A}}_y\rangle\bigg)\,\nonumber\\ 
&+ [1-N^{-1}]\bigg( q  \langle \mathcal{A}_x\rangle\,\langle\psi(Y)\dot{X}\rangle + q\,\langle \mathcal{A}_y\rangle\,\langle\chi^\prime(Y)\,\dot{Y}\rangle  \bigg)\nonumber\\
&- [1-N^{-1}]\epsilon_0(M_{\psi}\, \langle \mathcal{E}_x\rangle\,\langle \dot{\mathcal{A}}_x\rangle  +M_{\chi^\prime}\, \langle \mathcal{E}_y\rangle\,\langle \dot{\mathcal{A}}_y \rangle) - H(Z),
\end{align}
where the Hamiltonian ${H}$ is 
\begin{align}
H(Z) &=\frac{1}{2}\,m\,\langle V_x^2\rangle +\frac{1}{2}\,m\,\langle V_y^2\rangle  + \frac{1}{2}\,m\,\langle V_z^2\rangle\nonumber\\
&+\frac{1}{2}\epsilon_0\bigg(M_{\psi}\langle \mathcal{E}_x^2\rangle + M_{\chi^\prime}\langle \mathcal{E}_y^2\rangle\bigg)+\frac{1}{2}\mu_0^{-1}\bigg(M_{\psi^\prime}\langle \mathcal{A}_x^2\rangle\bigg)\nonumber\\
&+[1-N^{-1}]\frac{1}{2}\epsilon_0\bigg(M_\psi\langle\mathcal{E}_x\rangle^2 + M_{\chi^\prime}\langle\mathcal{E}_y\rangle^2\bigg)+[1-N^{-1}]\frac{1}{2}\mu_0^{-1}\bigg(M_{\psi^\prime}\langle \mathcal{A}_x\rangle^2\bigg).
\end{align}
Here we have set $\mathcal{W}(\bm{r}-\bm{X}) = \delta(\bm{r}-\bm{X})$ since there is no need to regularize the infinities associated with point-charge sources for Maxwell's equations, we have defined the real numbers
\begin{align*}
M_{\psi} = \int \psi(y)^2\,d^3\bm{r},\quad M_{\chi^\prime} = \int \chi^\prime(y)^2\,d^3\bm{r},\quad M_{\psi^\prime} = \int \psi^\prime(y)^2\,d^3\bm{r},
\end{align*}
and we have made use of the shorthand notation $\langle Q\rangle = \int Q(Z)\,\mathcal{F}_0(Z_0)\,dZ_0$.

Varying the truncated Lagrangian $L$ with respect to the fragment Lagrangian configuration map leads to the truncated Euler-Lagrange equations
\begin{gather}
m\,\dot{V}_x = q\,(\mathcal{E}_x+[1-N^{-1}]\langle \mathcal{E}_x\rangle)\,\psi(Y) - q\,V_y\,(\mathcal{A}_x + [1-N^{-1}]\langle \mathcal{A}_x\rangle)\,\psi^\prime(Y)\\
 m\,\dot{V}_y = q\,(\mathcal{E}_y+[1-N^{-1}]\langle \mathcal{E}_y\rangle )\chi^\prime(Y) + q\,V_x\,(\mathcal{A}_x + [1-N^{-1}]\langle \mathcal{A}_x\rangle)\,\psi^\prime(Y)\\
 m\,\dot{V}_z =0\\
 \epsilon_0\,M_\psi\,\dot{\mathcal{E}}_x + q\,\psi(Y)\,V_x =\mu_0^{-1}\,M_{\psi^\prime}\,\mathcal{A}_x,\quad \epsilon_0\,M_{\chi^\prime}\,\dot{\mathcal{E}}_y + q\,\chi^\prime(Y)\,V_y = 0\\
\dot{X} = V_x,\quad \dot{Y} = V_y,\quad \dot{Z} = V_z\\
\dot{\mathcal{A}}_x = - \mathcal{E}_x,\quad \dot{\mathcal{A}}_y = - \mathcal{E}_y.
\end{gather}
These equations comprise an evolution equation for the fragment Lagrangian configuration map $Z$. Since the single-fragment distribution function $\mathcal{F} $ is advected by $Z$, this implies
\begin{align}
\partial_t\mathcal{F} &+ \partial_x(V_x\,\mathcal{F}) + \partial_y(V_y\,\mathcal{F}) + \partial_z(V_z\,\mathcal{F})-\partial_{a_x}(e_x\,\mathcal{F}) - \partial_{a_y}(e_y\,\mathcal{F})\nonumber\\
& + \partial_{v_x}\bigg(\left\{ \frac{q}{m}\,(e_x+[1-N^{-1}]\langle e_x\rangle)\,\psi(y) -  \frac{q}{m}\,v_y\,(a_x + [1-N^{-1}]\langle a_x\rangle)\,\psi^\prime(y)\right\}\mathcal{F}\bigg)\nonumber\\
& + \partial_{v_y}\bigg(\left\{ \frac{q}{m}\,(e_y+[1-N^{-1}]\langle e_y\rangle)\,\chi^\prime(y) + \frac{q}{m}\,v_x\,(a_x + [1-N^{-1}]\langle a_x\rangle)\,\psi^\prime(y)\right\}\mathcal{F}\bigg)\nonumber\\
&+\partial_{e_x}\bigg(\left\{\frac{\mu_0^{-1}\,M_{\psi^\prime}}{\epsilon_0\,M_\psi}\,a_x - \frac{q}{\epsilon_0\,M_\psi}\psi(y)\,v_x\right\}\mathcal{F}\bigg) - \partial_{e_x}\bigg(\left\{\frac{q}{\epsilon_0\,M_{\chi^\prime}}\chi^\prime(y)\,v_y\right\}\mathcal{F}\bigg)=0.\label{trunc_fun_vlas}
\end{align}
The equation \eqref{trunc_fun_vlas} comprises a truncation of the fragment kinetic equation. The single unknown is $\mathcal{F}$, a scalar function on the $10$-dimensional space of tuples \[z=(x,y,z,a_x,a_y,v_x,v_y,v_z,e_x,e_y)\].

\section{Discussion\label{sec:discussion}}
It is useful to compare the fragment kinetic formalism presented here with the functional Schrodinger picture of quantum electrodynamics (QED). The Schrodinger wave functional $\Psi$ for scalar QED assigns a probability amplitude to each (single-time) field configuration $(\psi,\bm{a})$, where $\psi$ denotes the Klein-Gordon field, and $\bm{a}$ denotes the vector potential associated with $\psi$. The Wigner functional \cite{IBB_Morrison_1997} associated with $\Psi$ therefore defines a quasi-probability distribution on the field-theoretic phase space, $W(\psi,\pi,\bm{a},\bm{e})$, where $\pi$ is the conjugate momentum associated with $\psi$ and $\bm{e}$ is the electric field. In the formalism presented here, when the non-relativistic particles are replaced with Klein-Gordon fields the single-fragment distribution $\mathcal{F}$ becomes a genuine probability distribution on the same space. This similarity is no mistake. Just as the Vlasov distribution from classical mechanics arises as a semiclassical limit of the Wigner function from quantum mechanics, $\mathcal{F}$ is best thought of as a semiclassical limit of the Wigner functional $W$. It would be interesting to re-derive the fragment kinetic equation from the QED functional Schrodinger equation along these lines.

The fragment kinetic equation, even in the mean-field limit, incorporates the strongest classical correlations between particles and the electromagnetic field non-perturbatively, and therefore provides an attractive model for intense interactions between light and fully-ionized particles in aggressively burning HED plasmas. However, the model is not without its flaws. The mean-field limit of the fragment kinetic equation derived above does not account for collisions between charged particles. This shortcoming is conceptually straightforward to address using familiar statistical closure techniques \cite{Bonitz_2016} applied to the $N$-fragment BBGKY hierarchy, but represents a shortcoming nonetheless. The present formalism also makes no attempt whatsoever to incorporate partially-ionized atomic species, a type of matter surely present in burning HED plasmas. This presents a serious conceptual challenege since most atomic physics machinery available today does not apply in high-field environments. Finally it neglects quantum effects that are surely important, like exchange correlations, as well as quantum effects that may eventually become important when experimental radiation intensities reach sufficiently large values, like pair production. The Wigner functional formulation of QED, mentioned above, offers a promising approach to including some of these effects, at least perturbatively.

The usual theoretical apparatus for modeling interactions between thermal radiation and plasma describes the radiation field using the specific radiative intensity $I$ and the single-particle distribution function $f$. Each of these objects comprises a scalar function on a $6$-dimensional single-particle phase space. In contrast, the fragment kinetic equation describes both plasma and radiation using a single scalar function $\mathcal{F}$ on an infinite-dimensional fragment phase space. In practical computations, the single-fragment phase space would always be discretized using, e.g., truncated Fourier series. But even if only a few, say $16$, Fourier modes are retained, the dimension of the domain of the function $\mathcal{F}$ would far exceed the phase space dimensions encountered in the traditional approach involving $f$ and $I$. This indicates that only mesh-free methods offer a practical means to simulating the fragment kinetic equation. Future work will describe a Monte Carlo simulation framework for the mean-field fragment kinetic equation that extends the variational particle-in-cell technique \cite{Squire_PIC_2012,Xiao_2013,Shadwick_2014,Xiao_2015,Qin_nuc_2016,He_2016,Kraus_GEMPIC_pub_2017,Burby_FD_2017}.

Molecular dynamics is to the usual Vlasov equation what N-fragment dynamics is to the fragment kinetic equation. Therefore N-fragment simulations, in conjunction with data-driven methods for learning closures \cite{Bois_2022} or even representing collision operators \cite{Miller_2021}, provide a means for identifying conventional photon collision operators that capture strong-field effects. This is true in spite of the fact that the fragment formalism bears little superficial resemblance to the usual thermal radiative transport formalism. The key observation in this regard is that while the thermal radiative transport equation, together with particle kinetic equations, does not imply the fragment kinetic model, the single-fragment distribution implies a particular evolution for $I$ and $f$. For instance, the specific radiative intensity may be recovered from as the trace of the Wigner function of the two-point covariance for the electromagnetic field, as in \cite{Lessig_2013}. Provided data-driven methods can be developed that respect the fundamental conservation laws of energy and momentum, this avenue may be interesting to pursue as a way of incorporating high-field effects into extant radiation transport codes.

\bibliography{cumulative_bib_file.bib}



\end{document}